\documentclass[a4paper]{article}
\begin{document}

\title{Symmetry reduction and superintegrable Hamiltonian systems\footnote{Talk delivered at the Workshop Higher Symmetries in Physics, Madrid (Spain) 2008. To be published in Journal of Physics Conference Series}}
\author{M. A. Rodr\'{\i}guez\footnote{rodrigue@fis.ucm.es}, P. Tempesta\thanks{p.tempesta@fis.ucm.es} \\ Dept.~F\'{\i}sica Te\'orica II\\ Facultad de F\'{\i}sicas\\ Universidad Complutense de Madrid\\ 28040--Madrid, Spain \and P. Winternitz\thanks{wintern@crm.umontreal.ca} \\ Centre de recherches math\'ematiques et \\ D\'ept.~de math\'ematiques et de statistique \\ Universit\'e de Montr\'eal\\ Montr\'eal (QC) H3C 3J7 Canada}

\maketitle

\begin{abstract}
We construct complete sets of invariant quantities that are integrals
of motion for two Hamiltonian systems obtained through a reduction
procedure, thus proving that these systems are maximally superintegrable.
We also discuss the reduction method used in this article and its possible
generalization to other maximally superintegrable systems.
\end{abstract}

\section{Introduction}

Superintegrable Hamiltonian systems, both in classical and in quantum mechanics, have attracted a lot of attention in the last decades, due to their physically relevant properties. The existence of closed periodic bounded orbits for the classical models, and of accidental degeneracy of the energy levels for the quantum ones (removed by the existence of complete sets of commuting observables), are among the most remarkable consequences of the rich symmetry structure possessed by these systems.

Recent investigations lead to the construction of new superintegrable potentials, in flat spaces and curved spaces. As is well known, a maximally superintegrable system in a $N$--dimensional configuration space has $2N-1$ integrals of motion, although only sets of  $N$ integrals (chosen in several different ways among the $2N-1$ ones) are in involution. Among the classical examples of this type of Hamiltonians are well known systems: the Coulomb potential and the harmonic oscillator, whose dynamics is described by Bertrand's theorem \cite{Bertrand} (for a review see, for instance, \cite{TW03})

Superintegrable systems are rare and not easy to construct. Nevertheless, the approach we describe here offers a systematic way to produce superintegrable systems.

In this contribution, we are primarily interested in the reduction of superintegrable systems by using an adaptation of the Marsden--Weinstein reduction scheme \cite{MW} and in the study of the properties of the corresponding reduced systems. We will not analyze here in full generality and in all details the setting of this method. Instead, we will focus on two particular, but very interesting, cases. Reduction methods have been used as a common tool to derive new dynamic Hamiltonians starting from free systems, defined in the original manifold (see, for instance, \cite{CCM}). Instead, in our approach, we start with a dynamics that is not free, and study its reductions. The original potentials (Coulomb and harmonic oscillator) are already superintegrable, and their invariants are known.

The first system, i.e. the Coulomb potential plus centrifugal terms, was analyzed in \cite{VE08} by using an interesting \textit{ad hoc} approach. We will present here a different point of view, which can be easily generalized to other models. We apply our approach to this model and to that one obtained as a reduction of the classical anisotropic harmonic oscillator. We recall that these systems,  are also exactly solvable, i.e. their quantum spectrum can be computed by purely algebraic means \cite{TTW}. This paper is based on the work \cite{RTW}. For a treatment of the quantum case, see \cite{EV08}.

\section{A geometric approach to the construction of superintegrable systems}

In this section, we briefly review the geometric settings motivating our results. We work on a $2N-$dimensional symplectic manifold $(M, \omega)$. Recall that a momentum map for the action of a group $G$ on $M$, whose associated algebra is $\mathbf{g}$, is a map $J: M\rightarrow \mathbf{g}^{*}$ such that
\begin{equation}
d(\langle J,X\rangle)=i_{X}\omega,
\end{equation}
for all $X\in \mathbf{g}$. The previous formula reduces to $J_{X}=i_{X}\theta$ if there exist a Liouville one--form, i.e $\omega=d\theta$.
We also need the equivariance assumption
\begin{equation}
\langle  \mathrm{Ad}^{*}_g \,\xi,X \rangle  = \langle \xi, \mathrm{Ad}_{g^{-1}}X \rangle,
\end{equation}
where $\mathrm{Ad}^{*}_g$ denotes the coadjoint action of $G$ in $\mathbf{g^{*}}$.

Let us consider now an integrable system, whose configuration space is $M$. Then, there exist $2N$ functions in involution $I_1,...,I_{2N}$:
\begin{equation}
\{I_{i}, I_{j} \} = 0, \hspace{5mm} i,j = 1,...,2N,
\end{equation}
whose differentials $dI_j$ are independent at each point of an open dense submanifold $K$ of $M$. The reduction procedure can be sketched as follows. Due to the isomorphism between the algebra of smooth functions on $M$ and that of the associated Hamiltonian vector fields, the vector fields $X_1,...,X_{2N}$ corresponding to the set of integrals in involution are in involution as well: $[X_i,X_k]=0$. Then, we can construct their symplectic action $G=R^{2N}$ on the manifold. Let $J$ be the associated momentum map. The theorem of Marden and Weinstein ensures us that, if $\mu$ is a regular value of $J$, then $P_{\mu}:=J^{-1}/G$ is a symplectic manifold, whose dimension is equal to $\dim M-N=N$.

The same construction can be used to generate new superintegrable system as a reduction of a given superintegrable one. In the following sections, we apply this approach, very general,  to two physically interesting cases, like the Coulomb potential and the anisotropic oscillator. We point out that the integrals of the motion of the two reduced systems are obtained as a consequence of the reduction scheme. The corresponding Hamiltonian vector fields are still in involution, since the equivariance condition ensures that the isomorphism between algebras described above is preserved. Also, the functional independence of the reduced integrals is guaranteed, if we work on a sufficiently regular manifold. 
In our specific examples, the integrals $K_1,...,K_N$ responsible for the reduction process are the components of the angular momentum, and the corresponding momentum mapping is $J=K_1\times\cdots\times K_N$. Consequently, the group action is defined by the cartesian product of $N$ copies of $SO(2)$. Finally, our reduced space is
\begin{equation}
P_{\mu}=J^{-1}(\mu)/T^{N},
\end{equation}
where $T^{N}$ is the $N$--dimensional torus. 
As a general observation, the reduced Hamiltonians, as well as the integrals of motion, although keep memory of the original ones, are essentially different in their structure. In this sense, the procedure appears to be nontrivial, since it generates new dynamics from known ones, as will be transparent from the subsequent discussion.

\section{The Coulomb potential in even dimension and its reduction}
The Coulomb potential
\begin{equation}
H^{\mathrm C}_6=\frac{1}{2}\sum_{i=1}^{N} \hat{p}^2_i-\frac{\gamma}{\hat{r}},\quad \hat{r}^2=\sum_{i=1}^{N} \hat{x}_i^2
\end{equation}
is a maximally superintegrable system, with first integrals which can be derived from the angular momentum and the Laplace--Runge--Lenz tensors:
\begin{equation}
L_{ij}=\hat{x}_i\hat{p}_j-\hat{x}_j\hat{p}_i,\quad A_i=\sum_{j=1}^N \hat{p}_j L_{ij} - \gamma \frac{\hat{x}_i}{\hat{r}},\quad i,j=1,\ldots,N
\end{equation}
\begin{equation}
\{H_6^{\mathrm{C}},L_{ij}\}=0,\quad \{H_6^{\mathrm{C}},A_i\}=0
\end{equation}
 Although the whole procedure can be established in arbitrary even dimensions, for the sake of simplicity we will study in this work the six dimensional case, since the reduced system lives in a physical three dimensional space.

Due to the rotational symmetry of the problem, we can introduce the set of multipolar coordinates
\begin{eqnarray}
\hat{x}_1=x_1\cos x_1,\quad \hat{x}_2=x_1\sin x_1\nonumber\\
\hat{x}_3=x_2\cos x_2,\quad \hat{x}_4=x_2\sin x_2\nonumber\\
\hat{x}_5=x_3\cos x_3,\quad \hat{x}_6=x_3\sin x_3
\end{eqnarray}
and write the Hamiltonian as
\begin{equation}
H^{\mathrm C}_6=\frac{1}{2} p^2-\frac{\gamma}{r}+\frac{p_4^2}{2x_1^2}+\frac{p_5^2}{2x_2^2}+\frac{p_6^2}{2x_3^2},\quad r=\sqrt{x_1^2+x_2^2+x_3^2},\quad p=\sqrt{p_1^2+p_2^2+p_3^2}
\end{equation}
Since the variables $x_4$, $x_5$ and $x_6$ are cyclic, the corresponding momenta are constants and then we obtain the reduced Hamiltonian:
\begin{equation}
H^{\mathrm C}_3=\frac{1}{2}p^2-\frac{\gamma}{r}+\frac{k_1}{x_1^2}+\frac{k_2}{x_2^2}+\frac{k_3}{x_3^2}
\end{equation}

This Hamiltonian lives in the reduced manifold $P_{\mu}$ (see Section 2), where the group $G$ is the product of three copies of $SO(2)$. 

The Hamiltonian $H_3^{\mathrm C}$ is maximally superintegrable, that is we can find a set of five functionally independent first integrals (although not all of them in involution). Four of them can be easily obtained, being quadratic in the momenta \cite{MS67}, \cite{Ev90a}. However, the fifth one is not quadratic and was recently computed in \cite{VE08}. The method used in that work was inspired by the case when one of the barriers ($k_i$) vanishes. In this situation (for instance $k_3=0$), the quantity ($\vec{L}$ is the usual angular momentum in the three dimensional space):
\begin{equation}
p_2L_1-p_1L_2-2x_3\bigg(-\frac{\gamma}{2r}+\frac{k_1}{x_1^2}+\frac{k_2}{x_2^2}\bigg)
\end{equation}
is an integral. The argument in \cite{VE08} is that the fifth integral (which is not quadratic in the momenta) should reduce to this one in the limit $k_3\to 0$. This observation allows to calculate the integral we need to prove maximal superintegrability  (see \cite{VE08} for the details)

We will follow here a different approach, based on the theory of the reduction of superintegrable systems described above. The set of first integrals of the first Hamiltonian $H_6^{\mathrm{C}}$ can be written in the new coordinates $x_i$. However, not all of them will be independent of $x_4,x_5,x_6$, so they do not yield first integrals of the reduced system. Our approach consists in finding which of these quantities will satisfy this constraint. Equivalently, we will compute first order integrals of the Hamiltonian $H_6^{\mathrm{C}}$ which Poisson commute with the generators of the rotation group $L_{12}$, $L_{34}$ and $L_{56}$ (which are the generators of the subgroup of the rotation group which is used to construct the symmetry reduction). In fact, if we call
\begin{equation}
\tilde{L}=L|_{\mathrm{reduced}},\quad \tilde{A}=A|_{\mathrm{reduced}}
\end{equation}
the following reduced operators are constants
\begin{equation}
\tilde{L}_{12}=\sqrt{2k_1},\quad \tilde{L}_{34}=\sqrt{2k_2},\quad \tilde{L}_{56}=\sqrt{2k_3}
\end{equation}
In fact, as we have explained in Section 2, the reduced phase space is 
$$J^{-1} (\sqrt{2k_1},\sqrt{2k_2},\sqrt{2k_3})/T^3$$

We will define a set of reduced quantities, which do not depend on $x_4,x_5,x_6$ and then, are first order integrals of the reduced system:
\begin{eqnarray}
E&=&H \\
I_1&=& \tilde{L}_{13}^2+\tilde{L}_{14}^2+\tilde{L}_{23}^2+\tilde{L}_{34}^2\\
I_2&=& \tilde{L}_{15}^2+\tilde{L}_{16}^2+\tilde{L}_{25}^2+\tilde{L}_{26}^2\\
I_3&=& \tilde{L}_{35}^2+\tilde{L}_{36}^2+\tilde{L}_{45}^2+\tilde{L}_{46}^2
\end{eqnarray}
\begin{equation}
\{H_3^{\mathrm C},I_i\}=0,\quad i=1,2,3
\end{equation}
We need a fifth quantity which is not quadratic. We can construct it by using the Laplace--Runge-Lenz tensor (remark that these quantities commute with $L_{12}$, $L_{34}$ and $L_{56}$):
\begin{eqnarray}
T_1=& \tilde{A}_{1}^2+\tilde{A}_{2}^2\\
T_2=& \tilde{A}_{3}^2+\tilde{A}_{4}^2\\
T_3=& \tilde{A}_{5}^2+\tilde{A}_{6}^2
\end{eqnarray}
\begin{equation}
\{H,T_i\}=0,\quad i=1,2,3
\end{equation}
The seven first integrals cannot be functionally independent. In fact we only need a fourth order first integral. For instance, we can choose
\begin{equation}
T=T_1+T_2+T_3
\end{equation}
which can be written as:
\begin{equation}
T=\bigg(\frac{1}{2r}(2p^2r^2-(\vec{p}\,\vec{r})^2)+2r\bigg(-\frac{\gamma}{r}+\frac{k_1}{x^2}+\frac{k_2}{y^2}+\frac{k_3}{z^2}\bigg)\bigg)^2+\gamma\frac{(\vec{p}\,\vec{r})^2}{r}-\frac{(\vec{p}\,\vec{r})^4}{4r^2}
\end{equation}
(see \cite{VE08}).
We have constructed a set of five functionally independent first integrals (including the Hamiltonian) showing the maximal superintegrability of the system. The generalization of the above discussion to any (even) dimension is straightforward.


\section{The anisotropic harmonic oscillator in even dimension and its reduction}
As in the previous section, our results can be easily extended to any even dimension, but we shall restrict this work to the four dimensional case and its reduction, the 2--dimensional case. Here we review the results contained in \cite{RTW}.

The anisotropic oscillator in the two--dimensional case, both in classical and quantum mechanics, was discussed by Jauch and Hill \cite{JH40}, \cite{De63}, \cite{Il63}. The system
\begin{equation}
H^{\mathrm A}_N=\frac{1}{2}\sum_{i=1}^{N} \hat{p}^2_i+\frac{1}{2}\sum_{i=1}^N\omega_i^2\hat{x}_i^2
\end{equation}
is also known to be superintegrable (in any dimension), if the ratios
of the frequencies are rational. Let us consider a $N=4$ dimensional space and assume
\begin{equation}
\frac{\omega _{1}}{n_{1}}=\frac{\omega _{2}}{n_{1}}=\frac{\omega _{3}}{n_{2}}=\frac{\omega _{4}}{n_{2}}=\omega
\end{equation}
that is, consider the Hamiltonian
\begin{equation}\label{aniosc}
H^{\mathrm A}_4=\frac{1}{2}\sum_{i=1}^{4} \hat{p}^2_i+\frac{1}{2\omega^2}\bigg(n_1^2\sum_{i=1}^2\hat{x}_i^2+n_2^2\sum_{i=3}^4\hat{x}_i^2\bigg)
\end{equation}
Following \cite{JH40}, we define a set of (complex) invariants, with coordinates $z_{i},\bar{z}_{i}$, $i=1,\ldots,4$:
\begin{equation}
z_{j}=\hat{p}_{j}-i n_{j}\omega \hat{x}_{j},\qquad\overline{z}_{j}=\hat{p}_{j}+i n_{j}\omega
\hat{x}_{j}.  \label{II.2}
\end{equation}
It is easily checked that the expressions%
\begin{equation}\label{int}
c_{jk}=z_{j}^{n_{k}}\bar{z}_{k}^{n_{j}}  \label{II.3}
\end{equation}
provide integrals of motion. In particular, among these
integrals we have the angular momenta
\begin{equation}
L_{ik}=\hat{x}_{i}\hat{p}_{k}-\hat{x}_{k}\hat{p}_{i}  \label{II.4}
\end{equation}
(when $n_{i}=n_{k}$) and the tensor
\begin{equation}
T_{ik}=\hat{p}_{i}\hat{p}_{k}+n_{i} n_{k}  \omega^2  \hat{x}_{i}\hat{x}_{k}  \label{II.5}
\end{equation}
We will now study one of the several possible reductions of the anisotropic oscillator (\ref{aniosc}) and establish the superintegrability of the corresponding reduced system, computing, in an explicit way, three functionally independent constants of motion.

According to the scheme adopted for the Coulomb case, let us consider the following change of coordinates:
\begin{eqnarray}\label{polar}
&\hat{x}_1=x_{1}\cos x_{3},\quad \hat{x}_{2}=x_{1}\sin x_{3}\nonumber\\
&\hat{x}_{3}=x_{2}\cos x_{4},\quad \hat{x}_{4}=x_{2}\sin x_{4}.
\end{eqnarray}
The corresponding momenta read
\begin{eqnarray}
& \hat{p}_{1}=-p_{3}\frac{\sin x_{3}}{x_{1}}+p_{1}\cos x_{3},\quad \hat{p}
_{2}=p_{3}\frac{\cos x_{3}}{x_{1}}+p_{1}\sin x_{3}, \nonumber \\
& \hat{p}_{3}=-p_{4}\frac{\sin x_{4}}{x_{2}}+p_{2}\cos x_{4},\quad \hat{p}
_{4}=p_{4}\frac{\cos x_{4}}{x_{2}}+p_{2}\sin x_{4}.\label{polarmomenta}
\end{eqnarray}
and the system in the new coordinates is:
\begin{equation}
H_4^{\mathrm A}=\frac{1}{2}(p_1^2+p_2^2)+\frac{p_3^2}{2x_1^2}+\frac{p_4^2}{2x_2^2}+\frac{1}{2}\omega^2(n_1^2 x_1^2+n_2^2 x_2^2\big)
\end{equation}
Setting
\begin{equation}
p_3=\sqrt{k_1},\quad p_4=\sqrt{k_2}
\end{equation}
we get the reduced system:
\begin{equation}\label{redosc}
H_2^{\mathrm A}=\frac{1}{2}(p_1^2+p_2^2)+\frac{k_1^2}{2x_1^2}+\frac{k_2^2}{2x_2^2}+\frac{1}{2}\omega^2 \bigg(n_1^2 x_1^2+n_2^2 x_2^2\bigg)
\end{equation}
Notice the appearance of Rosochatius--type terms \cite{Rosochatius} in the reduced system.

As in he Coulomb case, this Hamiltonian is defined in a reduced manifold $P_{\mu}$ and the group used in this reduction is $SO(2)\times SO(2)$.

The integrals of the system (\ref{aniosc}), given in (\ref{int}), will be used to construct a set of integrals for the reduced system. Those that will survive the reduction  are the ones that are left invariant by the $SO(2)\times SO(2)$ rotations. The arguments are the same we used in the Coulomb potential, in fact, they must Poisson commute with
\begin{eqnarray}
L_{12}=& \frac{{\mathrm{i}}}{2n_1\omega}(z_{1}
\bar{z}_{2}-z_{2}\bar{z}_{1})=\hat{x}_{1}\hat{p}_{2}-\hat{x}_{2}\hat{p}_{1}, \nonumber \\
L_{34}=& \frac{{\mathrm{i}}}{2n_2\omega}(z_{3}\bar{z}_{4}-z_{4}\bar{z}_{3})=\hat{x}_{3}\hat{p}
_{4}-\hat{x}_{4}\hat{p}_{3} .\label{a}
\end{eqnarray}

The computation of these commuting quantities is not as straightforward as in the Coulomb potential, where we found the integrals by simple inspection. We will use $z$ coordinates to solve our problem. The Poisson bracket can be written in terms of the $z_{i}$ variables as
\begin{equation}
\{f(z_{i},\bar{z}_{i}),g(z_{i},\bar{z}_{i})\} = -2\mathrm{i} \omega \sum_{k=1}^{N} \sum_{j=2k-1}^{2k} n_k \bigg(\frac{\partial f}{\partial z_{j}}\frac{\partial g}{\partial \bar{z}_{j}}-
\frac{\partial f}{\partial \bar{z}_{j}}\frac{\partial g}{\partial z_{j}}\bigg)\label{Poisson}
\end{equation}
hence, functions of $z_k,\bar{z}_k$ Poisson commuting with $L_{12}$ and $L_{34}$ must satisfy
\begin{eqnarray}
z_{2}\partial _{z_{1}}f-z_{1}\partial _{z_{2}}f+\bar{z}_{2}\partial _{\bar{z}
_{1}}f-\bar{z}_{1}\partial _{\bar{z}_{2}}f  =0, \nonumber \\
z_{4}\partial _{z_{3}}f-z_{3}\partial _{z_{4}}f+\bar{z}_{4}\partial _{\bar{z}
_{3}}f-\bar{z}_{3}\partial _{\bar{z}_{4}}f =0 \label{atri}.
\end{eqnarray}
A basis for the corresponding $SO(2)\times SO(2) $ invariants is given by
\begin{eqnarray}
\xi_1  =z_{1}^{2}+z_{2}^{2},\quad \bar{\xi}_1=\bar{z}_{1}^{2}+\bar{z}_{2}^{2},\quad \eta_1= z_{1}\bar{z}_{1}+z_{2}\bar{z}_{2}, \nonumber \\
\xi_3  =z_{3}^{2}+z_{4}^{2},\quad \bar{\xi}_3=\bar{z}_{3}^{2}+\bar{z}_{4}^{2},\quad \eta_2= z_{3}\bar{z}_{3}+z_{4}\bar{z}_{4},  \label{inv}
\end{eqnarray}
These rotationally invariants will be constants of motion of our system if they satisfy:
\begin{equation}\label{posledni}
\{ H_4^{\mathrm{A}}, f(\xi_1,\bar{\xi}_1,\eta_1,\xi_3,\bar{\xi}_3,\eta_2) \}=0.
\end{equation}

The following expressions are solutions of equation (\ref{posledni})
\begin{eqnarray}
&E_1 =  \frac{1}{2} (|z_1|^2+|z_2|^2),
& E_2 =\frac{1}{2} (|z_3|^2+|z_4|^2), \nonumber \\
&Q_1  = ({z}_1^2+z_2^2)^{n_2} (\bar{z}_3^2+\bar{z}_4^2)^{n_1},\label{b}
& \bar{Q}_1 = (\bar{z}_1^2+\bar{z}_2^2)^{n_2} (z_3^2+z_4^2)^{n_1}, \\
&I_1 = (z_1^2+z_2^2)(\bar{z}_1^2+\bar{z}_2^2),  & I_2 = (z_3^2+z_4^2)(\bar{z}_3^2+\bar{z}_4^2).\nonumber
\end{eqnarray}
however, only five of these integrals are functionally independent.

We now perform the reduction for the integrals of motion using the change of variables (\ref{polar}), (\ref{polarmomenta}). The integrals (\ref{a}) reduce to constants $L_{12}=\sqrt{k_1}, \, L_{34}=\sqrt{k_2}$. The first four integrals in (\ref{b}) reduce to nontrivial integrals for the Hamiltonian (\ref{redosc}), namely
\begin{eqnarray}
E_1 &=&\frac{1}{2} p_1^2+ \frac{k_1}{2x_1^2}+\frac{1}{2} n_1^2 \omega^2 x_1^2, \nonumber \\
E_2 &=&\frac{1}{2} p_2^2+ \frac{k_2}{2x_2^2}+\frac{1}{2} n_2^2 \omega^2 x_2^2 \nonumber \\
Q_1 &= &( p_1^2+\frac{k_1}{x_1^2}-n_1^2 \omega^2 x_1^2 - 2 \mathrm{i} n_1 \omega p_1 x_1 )^{n_2}  ( p_2^2+\frac{k_2}{x_2^2}-n_2^2 \omega^2 x_2^2 + 2 \mathrm{i}  n_2 \omega p_2 x_2 )^{n_1},  \\
\bar{Q}_1 & = &( p_1^2+\frac{k_1}{x_1^2}-n_1^2 \omega^2 x_1^2 + 2 \mathrm{i}  n_1 \omega p_1 x_1 )^{n_2} ( p_2^2+\frac{k_2}{x_2^2}-n_2^2 \omega^2 x_2^2 - 2 \mathrm{i}  n_2 \omega p_2 x_2 )^{n_1}.\nonumber
\end{eqnarray}
The remaining two integrals in (\ref{b}) give nothing new
\begin{equation}
I_1 =4 (E_1^2-k_1 n_1^2 \omega^2), \quad  I_2 = 4 (E_2^2-k_2 n_2^2 \omega^2).
\end{equation}

Three functionally independent real integrals of motion of the system with Hamiltonian (\ref{aniosc})
can be chosen to be
\begin{equation}
\{ E_1,E_2,Q=\frac{1}{2}(Q_1+\bar{Q}_1 )\}.
\end{equation}
They are of order $2$, $2$ and $2(n_1+n_2) $ in the momenta, respectively. Their existence is the proof
of the maximal superintegrability of the considered system.

The integral of motion  $Q$ simplifies to give a second order one in two cases (that were known previously \cite{FMSUW}, \cite{Winternitz}).
They are
\begin{description}
\item{I)} $n_1=n_2=1$
\begin{equation}\label{d}
\frac{4 E_1 E_2 - Q}{2 \omega^2} = (p_1 x_2-p_2 x_1)^2+\frac{k_1 x_2^2}{x_1^2}+\frac{k_2 x_1^2}{x_2^2}.
\end{equation}
\item{II)} $n_1=1,n_2=2,k_2=0 $
\begin{equation}\label{e}
\left( \frac{8 E_1^2 E_2-Q}{8 \omega^2} -k_1E_2\right)^{1/2} = p_1(x_2 p_1-x_1 p_2)-\omega^2 x_1^2 x_2+k_1 \frac{x_2}{x_1^2}
\end{equation}
\end{description}
The integrals (\ref{d}) and (\ref{e}) are responsible for the separation of variables in polar and parabolic coordinates, respectively.
The integrals $\{ E_1, E_2 \} $ are responsible for the separation in cartesian coordinates.

\section{Conclusions}

From the previous examples it is clear that it would be very interesting to construct, in a systematic way, transformations mapping a superintegrable system into another system, that is also superintegrable, and defined in a reduced phase space. The role of higher order groups of transformations generated by the flow associated to integrals that are polynomials in the momenta remains to be fully investigated. 

We also mention that an algebraic approach to superintegrable systems, independent of the geometry of the phase space, and based on a suitable nilpotency condition, is presently under investigation \cite{IMRT}.

A quantum mechanical version of this reduction procedure is also to be understood.


\section*{Acknowledgements} The research of P.W. was partly supported by NSERC of Canada. The research of M.A.R. and P.T. was supported by MICINN (Spain) under grant no. FIS2008-00752 and Universidad Complutense and Comunidad de Madrid under grant no. GR74/07-910556.


\section*{References}
\begin{thebibliography}{99}



\bibitem{AM78} Abraham R, Marsden J E 1978 {\it Foundations of Mechanics} (2nd ed., The Benjamin/Cummings Publishing Company, Reading).

\bibitem{Bertrand} J. Bertrand, Comptes Rendus. Acad. Sci. Paris, \textbf{77}, 849 (1873).

\bibitem{CCM} Cari\~{n}ena J F, Clemente--Gallardo J and Marmo G 2007 {\it Int. J. Geom. Methods Mod. Phys.} {\bf 4} 1363

\bibitem{De63} Demkov Yu N 1963 {\it Sov. Phys. JETP} {\bf 17} 1349.

\bibitem{Ev90a} Evans N W 1990 Phys. Rev. A {\bf 41}, 5666.

\bibitem{EV08} Evans N W and Verrier P E 2008 J. Math. Phys. {\bf 49} 092902.

\bibitem{FMSUW} Fri\v{s} J, Mandrosov V, Smorodinsky Ya A, Uhli\v{r} M and Winternitz P 1965 Phys. Lett. {\bf 16} 354.

\bibitem{IMRT} Ibort, A, Marmo G, Rodr\'{\i}guez M A  and Tempesta P, \textit{Nilpotent integrability and algebraic reductions in classical mechanics}, in preparation.

\bibitem{Il63} Ilkaeva L A 1963 {\it Vestnik LGU} {\bf 22} 56 (in Russian).

\bibitem{JH40} Jauch J and Hill E 1940 Phys. Rev. {\bf 57} 641.

\bibitem{MS67} Makarov A A, Smorodinsky J A, Valiev Kh and Winternitz P 1967 Nuovo Cimento A {\bf 52} 1061.

\bibitem{MW} Marsden J E and A. Weinstein 1974 {\it Rep. Math. Phys.} {\bf 5} 121.

\bibitem{MF} Mischenko A S and A. T. Fomenko 1978 {\it Funct. Anal. Appl.} {\bf 12} 133

\bibitem{Nekh} Nekhoroshev N N 1972 {\it Trans. Moscow Math. Soc.} {\bf 26} 180

\bibitem{RTW} Rodr\'{\i}guez M A, Tempesta P and Winternitz P 2008 Phys. Rev. E {\bf 78} 046608.

\bibitem{Rosochatius} Rosochatius E 1877 {\it Dissertation} (Gottingen, Gebr. Unger, Berlin)

\bibitem{TW03} Tempesta P, Winternitz P, Harnad J, Miller W Jr, Pogosyan G, Rodr\'{\i}guez M A (eds) 2004 {\it Superintegrability in Classical and Quantum Systems}, (Montr\'{e}al, CRM Proceedings and Lecture Notes, AMS vol. 37).

\bibitem{TTW} Tempesta P, Turbiner A and Winternitz P 2001 J. Math. Phys. {\bf 42} 4248

\bibitem{VE08} Verrier P E  and Evans N W  2008 J. Math. Phys. {\bf 49} 022902.

\bibitem{Winternitz} Winternitz P, Smorodinsky Ya, Uhli\v{r} M and Fri\v{s} J 1966 {\it Yad. Fiz.} {\bf 4} 625; 1967 {\it Sov. J. Nucl. Phys.} {\bf 4} 444.

%






%

%



%



%







%



%


\end{thebibliography}
\end{document}